\documentclass{PoS}

\usepackage{amsmath}
\usepackage{amsthm}
\usepackage{amsfonts}

\title{Heavy-light hadrons and their excitations}

\ShortTitle{Heavy-light hadrons and their excitations}

\author{Tommy Burch$^a$, \speaker{Christian Hagen}$^b$, Christian B. Lang$^c$, Markus Limmer$^c$ and Andreas Sch\"afer$^{b,d}$\\
\llap{$^a$}
Department of Physics, 
University of Utah,\\
Salt Lake City, UT 84112, U.S.A.\\
\llap{$^b$}
Institut f\"ur Theoretische Physik,
Universit\"at Regensburg,\\
D-93040 Regensburg, Germany\\
\llap{$^c$}
Institut f\"ur Physik, FB Theoretische Physik,
Karl-Franzens-Universit\"at Graz,\\
A-8010 Graz, Austria\\
\llap{$^d$}
Yukawa Institute of Theoretical Physics,
Kyoto University,\\
Kyoto 606,8502, Japan\\

E-mail:\\
\email{tburch@physics.utah.edu},
\email{christian.hagen@physik.uni-regensburg.de},
\email{christian.lang@uni-graz.at},
\email{markus.limmer@uni-graz.at},
\email{andreas.schaefer@physik.uni-regensburg.de}

}

\abstract{
We study the excitations of hadrons containing a single heavy quark. We present meson and baryon mass splittings and ratios of meson decay constants resulting from quenched and dynamical two-flavor configurations. Light quarks are simulated using the Chirally Improved (CI) lattice Dirac operator. The heavy quark is approximated by a static propagator, appropriate for the $b$ quark on our lattices ($1/a \sim 1-2$ GeV). We also include some preliminary calculations of the heavy-quark kinetic corrections to the states.
}

\FullConference{The XXVI International Symposium on Lattice Field Theory\\
		 July 14-19 2008\\
		 Williamsburg, Virginia, USA}

\begin{document}

\section{Introduction}

The heavy quark is approximated by a static propagator, while for the light quark 
propagator we use estimated all-to-all propagators. To improve these estimates we 
use the so-called Domain Decomposition Improvement \cite{DDI}.
In these proceedings we describe only briefly the methods which we use to extract mass splittings, ratios of decay constant, and kinetic corrections for static-light hadrons. Also we show only a subset of our results. A complete discussion of the methods and our results is presented in Ref. \cite{Burch:2008qx}.

\section{Methodology}

\subsection{Masses}

The masses of ground and excited states are extracted with the variational method \cite{varmethod}. For that purpose we construct the different interpolating fields for the states by using different ``wavefunctions'' for the light-quark source and sink. The cross-correlation matrix $C(t)$, which we can compute from them, is then inserted into the generalized eigenvalue problem:
\begin{align}
C(t) \; \vec{\psi}^{(\alpha)} = \lambda^{(\alpha)}(t,t_0) C(t_0) \; \vec{\psi}^{(\alpha)} \; .
\label{math:geneval}
\end{align}
For sufficiently large $t\gg t_0$, the eigenvalues are 
\begin{align}
\lambda^{(\alpha)}(t,t_0) = c^{(\alpha)} e^{-(t-t_0) M^{(\alpha)}} \left[1 + O(e^{-(t-t_0) 
\Delta^{(\alpha)}})\right] \; ,
\end{align}
where $\Delta^{(\alpha)}$ is the energy difference to the closest state. However, due to the static quark propagator only differences to the so extracted lattice energies $M^{(\alpha)}$ are physically meaningful. Thus, we only report mass splittings in the following.

\subsection{Couplings}
\label{sect:couplings}

Using the eigenvectors obtained by solving Eq. (\ref{math:geneval}), we can construct and fit the ratio
\begin{align}
R(t)_i^{(\alpha)} = \frac{\left| \sum_j C(t)_{ij} \psi_j^{(\alpha)} \right|^2 }{\sum_i \sum_j \psi_i^{(\alpha)*} C(t)_{ij} \psi_j^{(\alpha)}}
\approx v_i^{(\alpha)} v_i^{(\alpha)*} \, e^{-t E^{(\alpha)}} \; .
\label{math:couplingratio}
\end{align}
which allows us to determine the couplings $v$. Ratios of different couplings to the same mass eigenstate \cite{ExcAmp} are even more straightforward:
\begin{align}
\frac{v_i^{(\alpha)}}{v_k^{(\alpha)}} \approx \frac{\sum_j C(t)_{ij} \psi_j^{(\alpha)}}{\sum_l C(t)_{kl} \psi_l^{(\alpha)}} \; .
\end{align}
Then, for example, the coupling of the local vector operator ($O_i=\bar{q}\gamma_iQ$, where $q$ and $Q$ denote the light and heavy quark, respectively) can be related to the pseudoscalar decay constants via
\begin{align}
f_{PS}^{(\alpha)} = \sqrt{\frac{2}{M^{(\alpha)}}}  \left( v_i^{(\alpha)} + \mathcal{O}(k^{(\alpha)2})/m_Q \right) \; .
\end{align}
In order to cancel renormalization constants and matching coefficients between HQET and QCD, we deal with the ratios of of decay constants. So, for example, the ratio $f_{B_s'}/f_{B_s}$, may be extracted from the $m_q=m_s$ point of
\begin{align}
\frac{f_{PS}^{(2)}}{f_{PS}^{(1)}} = \frac{v_i^{(2)}}{v_i^{(1)}} 
\sqrt{\frac{M_{B_s^{(*)}}}{M_{B_s^{(*)}}+(E^{(2)}-E^{(1)})}} \; + \, {\cal O}\left( k_s^{(2)2}/m_b^{} \, , \, 
k_s^{(1)2}/m_b^{} \right) \; .
\end{align}
where $M_{B_s^{(*)}} = 5400$ MeV and the $v_i^{(\alpha)}$ and $E^{(\alpha)}$ come from fits to Eq.\ (\ref{math:couplingratio}) for $\alpha = 1$ and 2.

\subsection{Kinetic corrections}
\label{sect:kincorr}

The $\mathcal{O}(1/m_Q)$ kinetic corrections to the static approximation can be incorporated into the simulations in form of lattice three-point functions
\begin{align}
T(t,t')_{ij} = \langle \, 0 \, | \, \bar q O_i Q(t) \; \bar Q \vec D^2 Q(t') \; \bar Q O_j^\dagger q(0) \, | \, 0 \, \rangle \; ,
\label{math:threept}
\end{align}
with the current insertion $\bar Q \vec D^2 Q(t')$, where $\vec D^2$ is the lattice-discretized covariant Laplacian. To obtain the corrections not only for the lowest lying state but also for the excitations, we consider two separate variational problems:
\begin{align}
\nonumber  C(t-t')_{ij} &= \langle \, 0 \, | \, \bar q O_i Q(t) \; \bar Q O_j^\dagger q(t') \, | \, 0 \, \rangle, \\
C(t')_{ij} &= \langle \, 0 \, | \, \bar q O_i Q(t') \; \bar Q O_j^\dagger q(0) \, | \, 0 \, \rangle.
\end{align}
Solving the two corresponding generalized eigenvalue equations,
\begin{align}
\nonumber  {\bf C}(t-t') \, \vec{\psi}^{(\alpha)} &= \lambda(t-t',t_0'-t')^{(\alpha)} \, {\bf C}(t_0'-t') \, \vec{\psi}^{(\alpha)},  \\
{\bf C}(t') \, \vec{\phi}^{(\beta)} &= \lambda(t',t_0)^{(\beta)} \, {\bf C}(t_0) \, \vec{\phi}^{(\beta)}.
\end{align}
then gives sets of eigenvectors, which can be used to project the states of interest in Eq. (\ref{math:threept}). To cancel exponentials and some
overlap factors one may form ratios:
\begin{align}
  R_{kl}^{(\alpha,\beta)} =
  \frac{ \sum_i \sum_j \psi_i^{(\alpha)*} T(t,t')_{ij} \phi_j^{(\beta)} }
  { \sum_a \psi_a^{(\alpha)*} C(t-t')_{ak} \cdot
    \sum_b C(t')_{lb} \phi_b^{(\beta)} } \approx \frac{\varepsilon^{(\alpha,\beta)}}{v_k^{(\alpha)*} v_l^{(\beta)}},
\end{align}
where the $\varepsilon^{(\alpha,\beta)}$ represent the matrix elements relevant for the kinetic corrections.

\section{Simulation details}

Our calculations are performed using CI fermions \cite{CIferm} for the light valence quarks. 
For the quenched simulations we work with three lattice sizes with approximately the same spatial volume 
of 2.4 fm and lattice spacings reaching from 0.20-0.12 fm. A necessary step for the CI operator 
is to smear the configurations. For the quenched lattices one step of HYP-smearing is 
applied. Another part of our simulations are done on two sets of lattices with $N_f=2$ 
dynamical CI-fermions \cite{DynCI}. Here one step of stout smearing is 
used. In all cases we use the one-loop improved L\"uscher-Weisz gauge action \cite{LWgauge}.

In order to increase the number of basis operator for the variational method, we create quark sources and sinks with a number ``shapes'' by using different covariant spatial smearings.

\section{Results for mass splittings}

\begin{figure}
\begin{center}
\resizebox{0.7\textwidth}{0.31\textwidth}{
\includegraphics[clip]{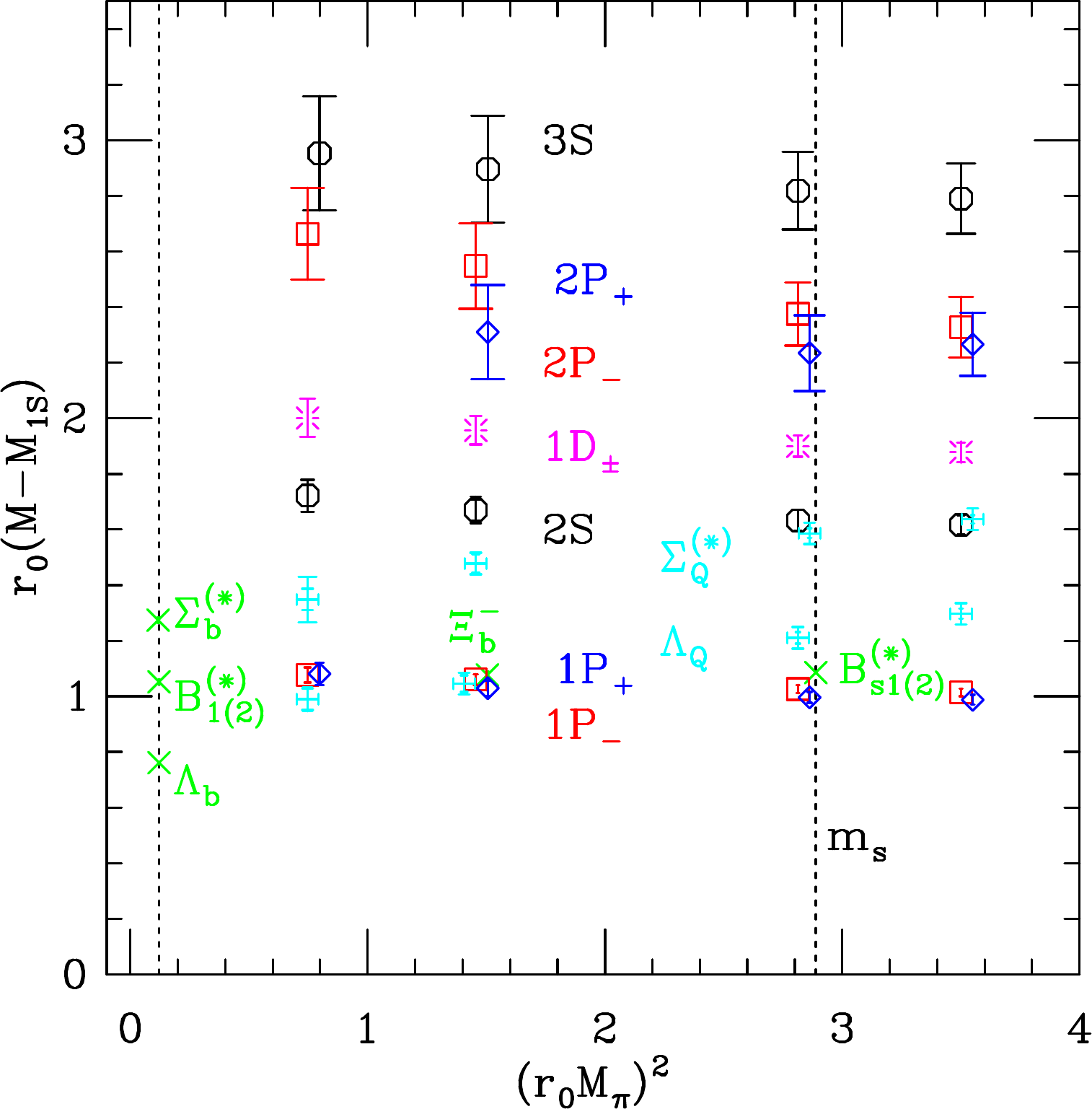} \hspace{1cm} \includegraphics[clip]{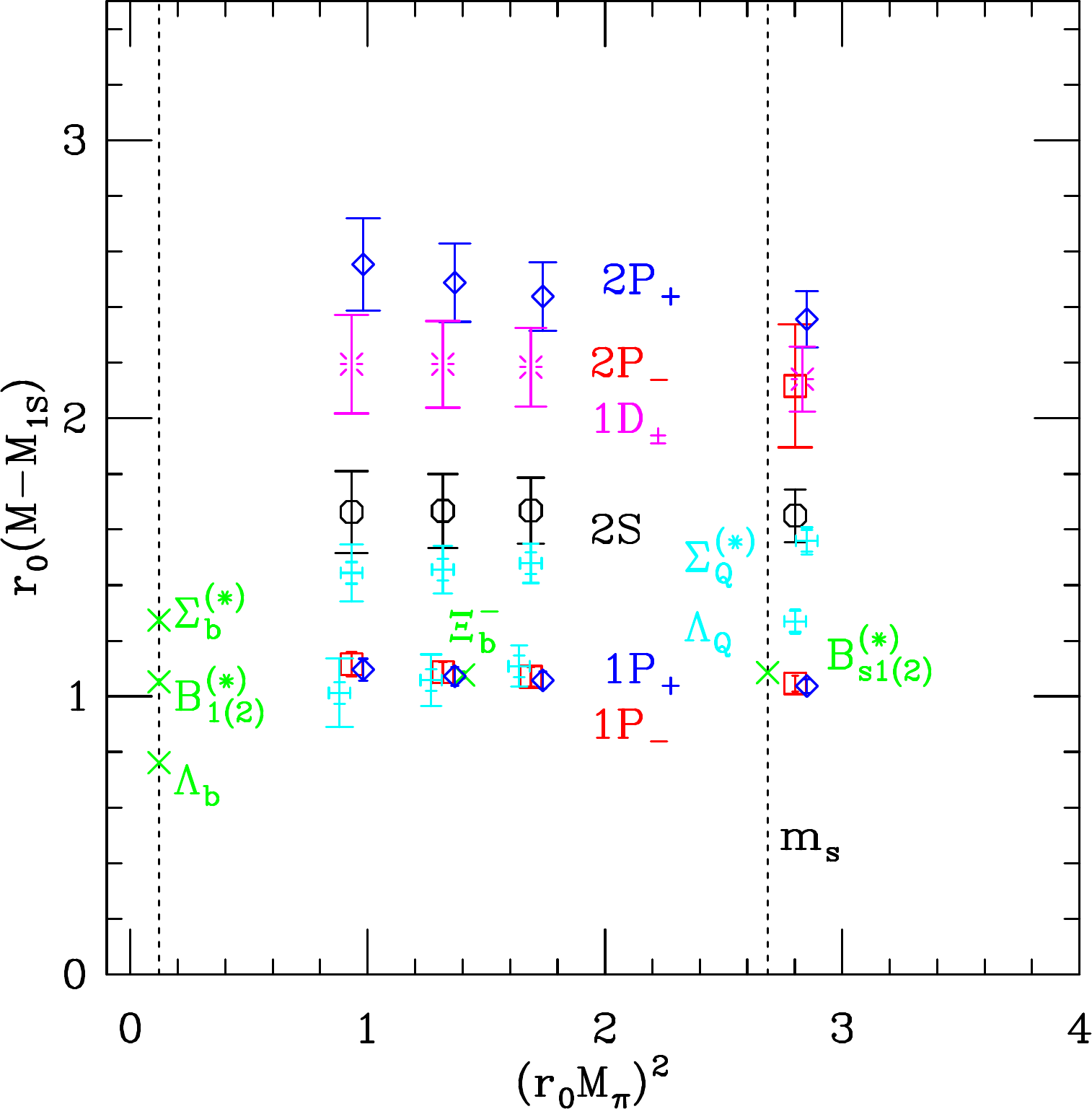}
}
\end{center}
\caption{
Meson and baryon masses, relative to the meson ground state $M_{1S}=M(B_q^{(*)})$, as a function 
of $M_\pi^2$ ($\propto m_q$) on the $16^3\times32$, $\beta=7.90$ quenched lattice (left plot) and 
the $16^3\times32$, $\beta=4.65$ dynamical lattice (right plot). All masses result from fits to the 
eigenvalues of the complete $4\times4$ basis. Circles represent the $S$ states; squares, the $P_-$ 
states; diamonds, the $P_+$ states; bursts, the $1D_\pm$; pluses, the baryons $\Lambda_Q$ and 
$\Sigma_Q^{(*)}$; and crosses, the experimental results \cite{expres}
(using $r_0=0.49$ fm).
}
\label{fig:16lattices}
\end{figure}

In the following we present a selection of our results for the static-light hadrons mass splittings. Figure 
\ref{fig:16lattices} shows exemplarily the results obtained on the $16^3\times32$, $\beta=7.90$ 
quenched lattice (left plot) and the $16^3\times32$, $\beta=4.65$ dynamical lattice (right plot). 
The vertical dashed lines indicate the physical pion mass and the pion mass which corresponds to 
the strange quark mass on these lattices. The latter is set via the splitting 
$M_{1S_s} - M_{1S_{ud}} = 76.9$ MeV, which is the $1/M_{H^{(*)}} \rightarrow 0$ linear 
extrapolation of the experimental values $M_{B_s^{(*)}}-M_{B^{(*)}}=86.8$ MeV and 
$M_{D_s^{(*)}}-M_{D^{(*)}}=103.5$ MeV. On both lattices we can extract a number of excited states, 
including a 3S state in the quenched case. The results on the dynamical lattices are consistent 
with those on the quenched configurations but have much larger error. This is most likely due
to the additional fluctuations which are added by the sea quarks.

\begin{figure}
\begin{center}
\resizebox{0.7\textwidth}{0.31\textwidth}{
\includegraphics[clip]{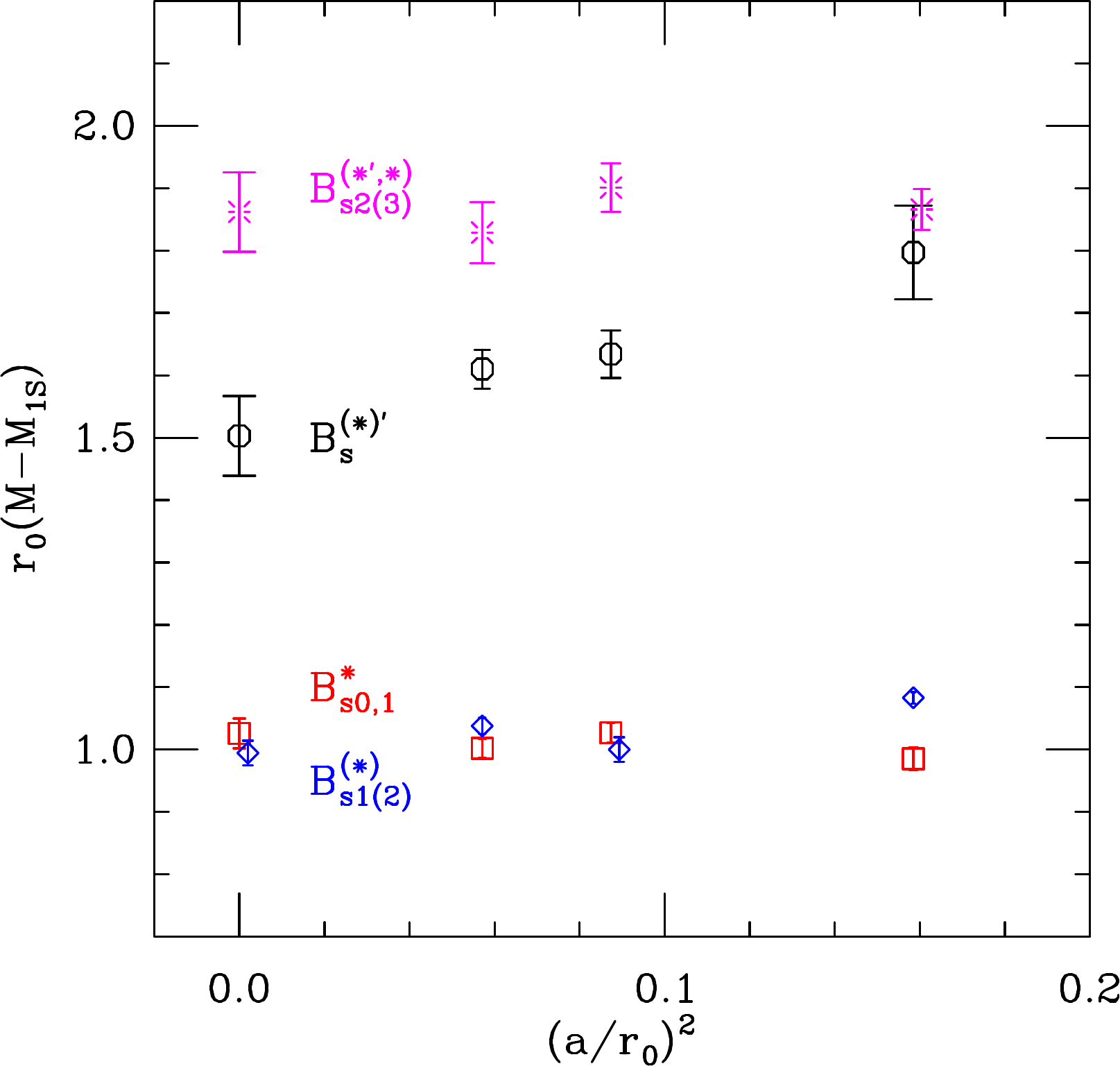} \hspace{1cm} \includegraphics[clip]{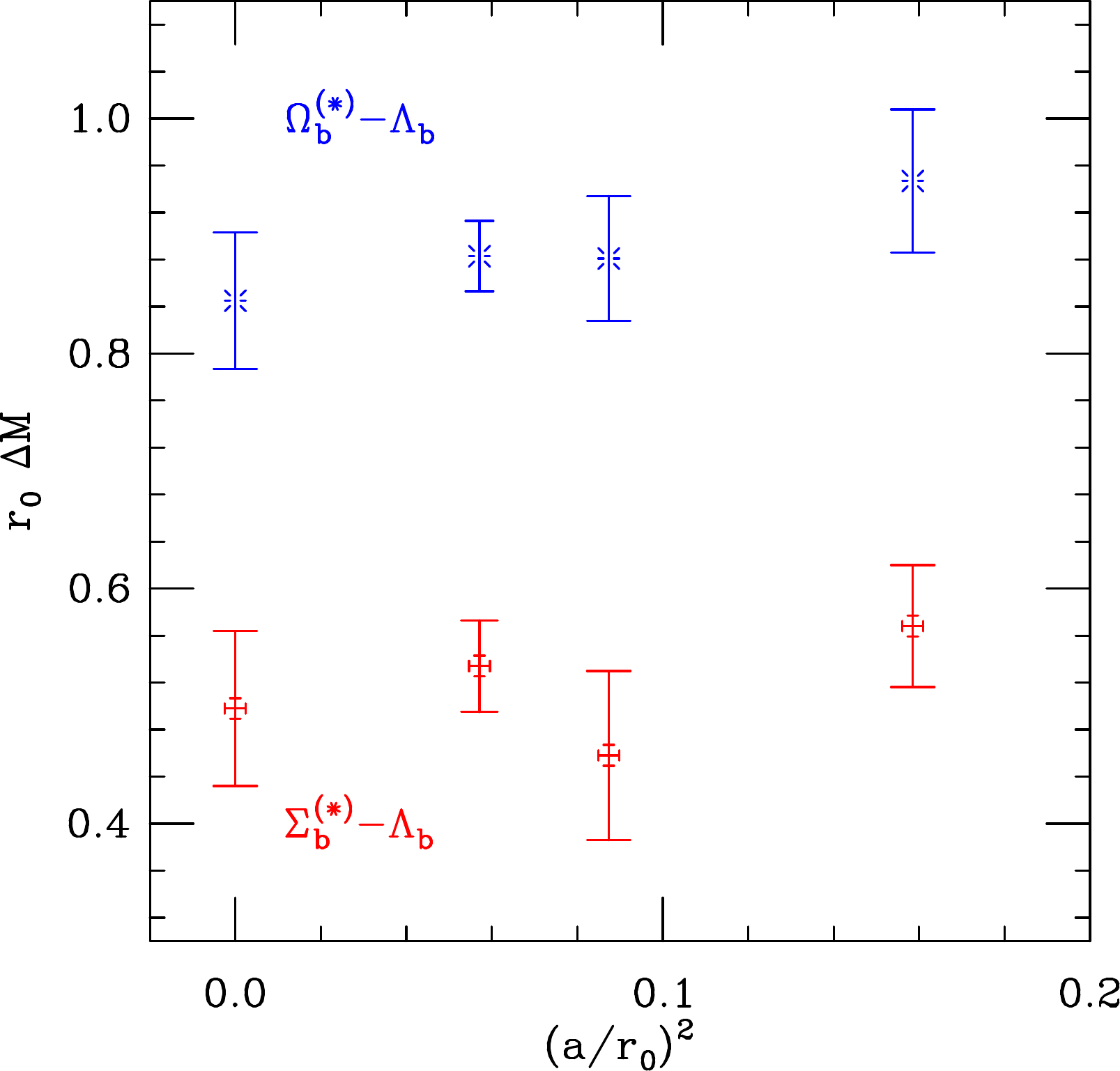}
}
\end{center}
\caption{
Quenched continuum extrapolations of excited $B_s$ meson masses, relative to $B_s^{(*)}$ (left plot) and quenched continuum extrapolations of the $\Sigma_b^{(*)}-\Lambda_b$ and $\Omega_b^{(*)}-\Lambda_b$ baryon mass differences (right plot).
}
\label{fig:contextrap}
\end{figure}

On the left hand side of Figure \ref{fig:contextrap}, we show continuum extrapolations for our 
quenched results for mass splitting of the $B_s$ states. While the results on the two finer lattice
agree with each other we find a significant difference for the coarsest lattice, especially for
the 2S state. This might be a hint for possible discretization errors on that lattice. Nevertheless, for
the continuum extrapolation we include also the results on the coarsest lattice. The right plot in 
Figure \ref{fig:contextrap} shows the corresponding extrapolation for mass differences of some of 
the baryons alone.

\begin{table}
\begin{center}
\footnotesize
\begin{tabular}[t]{cccc}
\hline \hline
difference & $N_f=0$ & $N_f=2$ & experiment \\
 & $a\rightarrow0$ & $a=0.156(3)$ fm & \\
\hline
$B_{1(2)}^{(*)}-B^{(*)}$ & 423(13)(9) & 446(17)(9) & 423(4) \\
$B_{s1(2)}^{(*)}-B_s^{(*)}$ & 400(8)(8) & 417(10)(9) & 436(1) \\
\hline
$\Lambda_b-B^{(*)}$ & 415(23)(8) & 358(55)(7) & 306(2) \\
$\Sigma_b^{(*)}-B^{(*)}$ & 604(16)(12) & 555(47)(11) & 512(4) \\
$\Xi_b-B^{(*)}$ & 466(17)(10) & 426(37)(9) & 476(5) \\
\hline
$\Sigma_b^{(*)}-\Lambda_b$ & 200(27)(4) & 195(72)(4) & 206(4) \\
$\Xi_b-\Lambda_b$ & 95(17)(2) & 111(37)(2) & 170(5) \\
\hline \hline
\end{tabular}
\hspace{1mm}
\begin{tabular}[t]{ccc}
\hline \hline
difference & $N_f=0$ & $N_f=2$ \\
 & $a\rightarrow0$ & $a=0.156(3)$ fm \\
\hline
$B^{(*)'}-B^{(*)}$ & 612(31)(13) & 674(66)(14) \\
$B_s^{(*)'}-B_s^{(*)}$ & 604(26)(12) & 664(39)(13) \\
$B_{0,1}^*-B^{(*)}$ & 435(15)(9) & 454(19)(9) \\
$B_{s0,1}^*-B_s^{(*)}$ & 412(10)(8) & 421(12)(9) \\
\hline
$\Omega_b^{(*)}-B_s^{(*)}$ & 683(9)(14) & 624(21)(13) \\
\hline
$\Omega_b^{(*)}-\Lambda_b$ & 340(23)(7) & 342(55)(7) \\
$\Xi_b^{(',*)}-\Lambda_b$ & 272(23)(6) & 269(55)(5) \\
$\Xi_b^{(',*)}-\Xi_b$ & 173(20)(4) & 158(50)(3) \\
\hline \hline
\end{tabular}
\end{center}
\caption{
Summary of mass differences (in MeV). When possible, we compare our results to experimental numbers
\cite{expres}. Our values are given using $r_0=0.49(1)$ fm.
The first error of our results is the statistical error while the second one is
a systematic error coming from the uncertainty of $r_0$.
}
\label{tab:splittings}
\end{table}

Table \ref{tab:splittings} summarizes the results of our simulations in physical units and compares 
(where possible) with experimental values.

\section{Results for decay constants}

\begin{figure}
\begin{center}
\resizebox{0.7\textwidth}{0.31\textwidth}{
\includegraphics[clip]{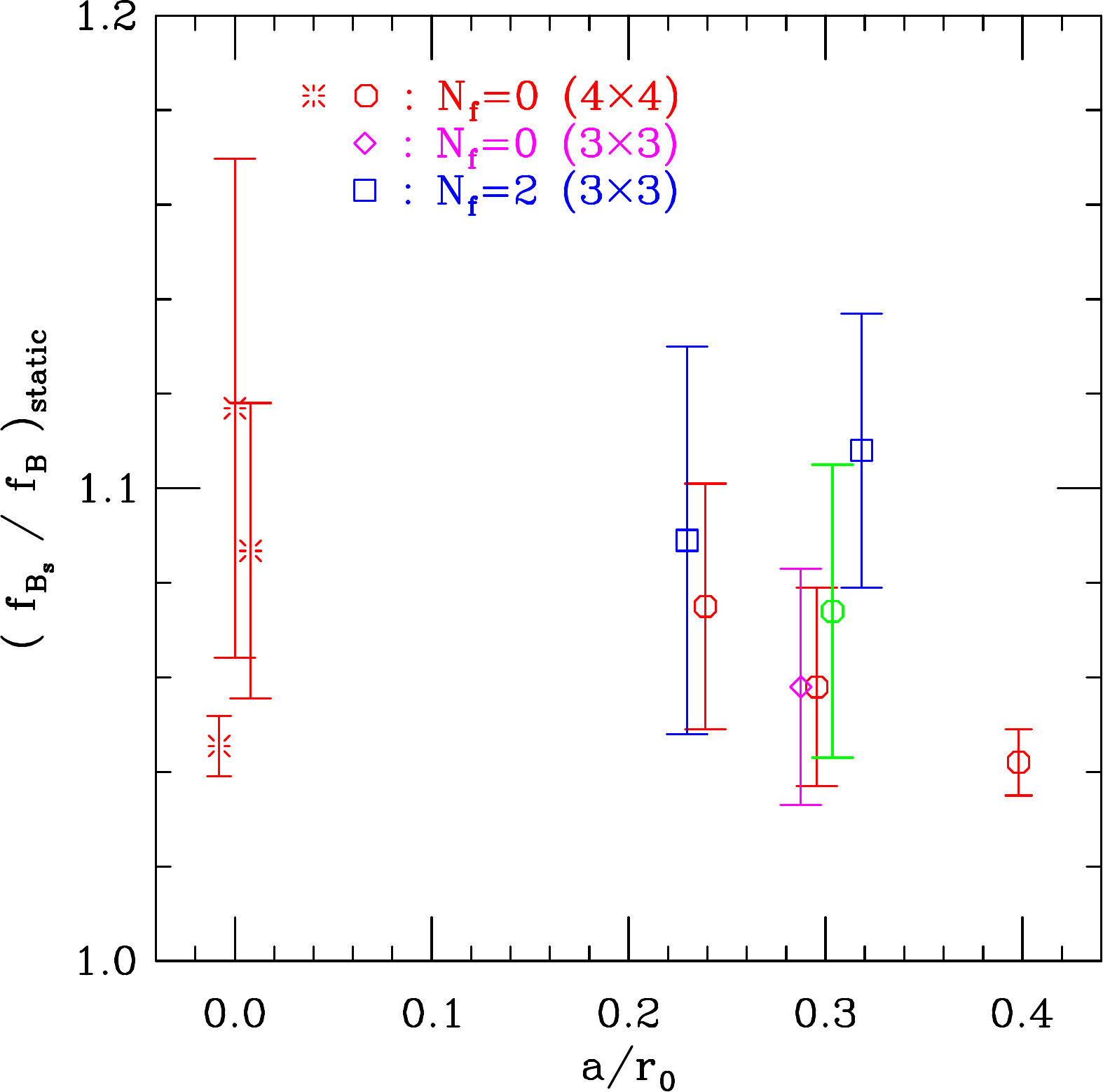} \hspace{1cm}\includegraphics[clip]{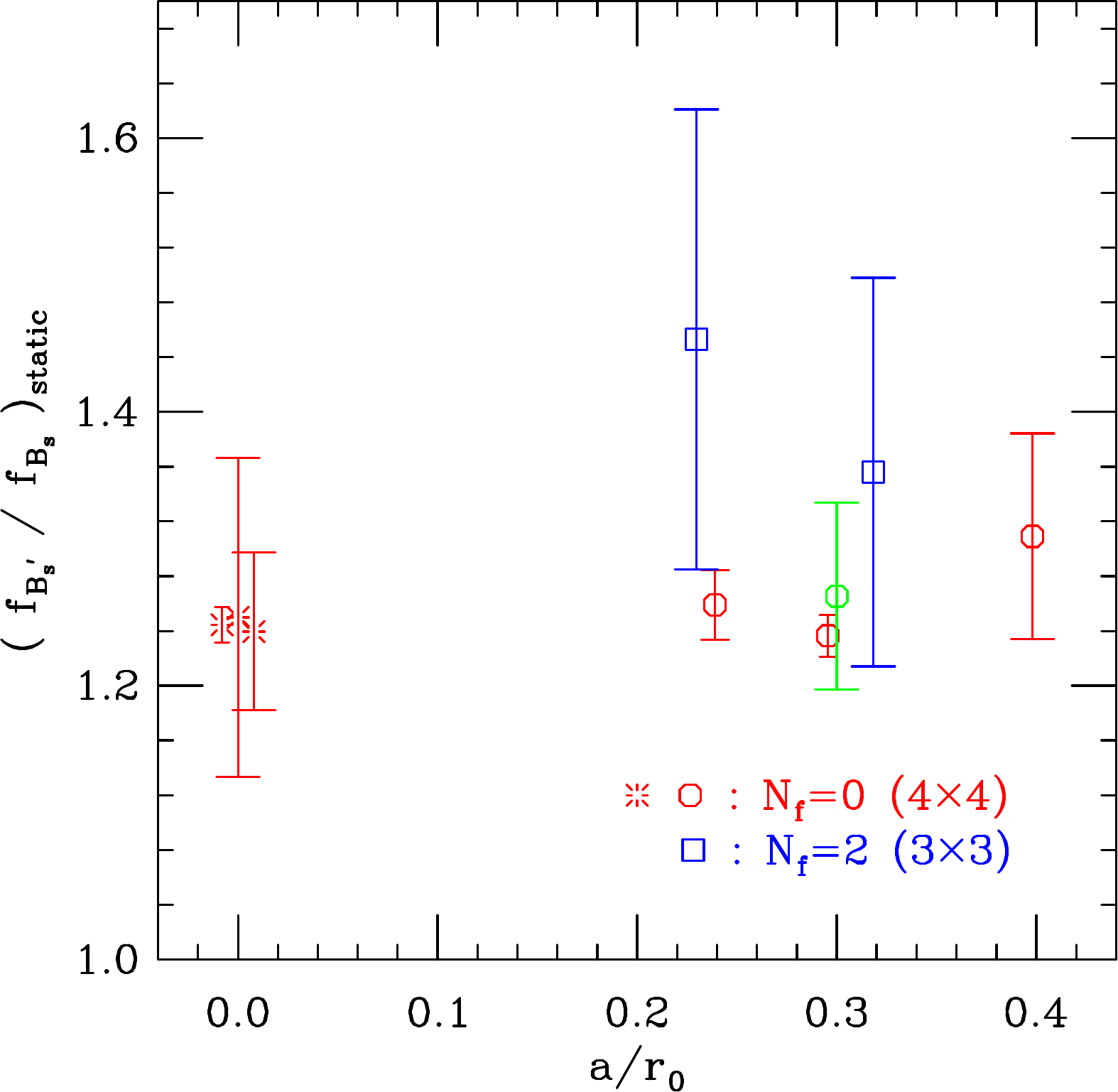}
}
\end{center}
\caption{
Ratios of meson decay constants $(f_{B_s}/f_B)_{static}$ (left plot) and 
$(f_{B_s'}/f_{B_s})_{static}$ (right plot) as a function of lattice spacing. 
The three quenched, continuum extrapolations correspond to (from left to 
right): a constant fit, a fit linear in $a$, a fit linear in $a^2$. 
The green circle at $a/r_0 \approx 0.3$ corresponds to the adjacent quenched 
result, but with thin links for the static quark.
}
\label{fig:decayratio}
\end{figure}

In this section we present first results for the ratios of meson decay constants 
using the method described in Section \ref{sect:couplings}. Figure \ref{fig:decayratio}
shows our results for the ratios of meson decay constants $(f_{B_s}/f_B)_{static}$ 
(left plot) and $(f_{B_s'}/f_{B_s})_{static}$ (right plot) as a function of lattice 
spacing. For the continuum extrapolation of the quenched results we try different 
fit functions. Apart from these ratios we also extract values for $(f_{B_s'}/f_{B'})_{static}$
ratios of couplings of excited states. The numerical results for the ratios are summarized in 
Table \ref{tab:decayratio}.

\begin{table}
\begin{center}
\footnotesize
\begin{tabular}{cccc}
\hline \hline
$\beta$ & $(f_{B_s}/f_B)_{static}$ & $(f_{B_s'}/f_{B_s})_{static}$ & $(f_{B_s'}/f_{B'})_{static}$ \\
\hline
7.57 & 1.042(7) & 1.309(75) & 0.976(142) \\
7.90 & 1.058(21) & 1.237(15) & 0.996(59) \\
8.15 & 1.075(26) & 1.259(25) & 0.972(61) \\
$\infty$ & 1.087(31) & 1.240(58) & 0.972(123) \\
4.65 & 1.108(29) & 1.356(142) & 1.089(259) \\
5.20 & 1.089(41) & 1.453(168) & 1.026(128) \\
\hline \hline
\end{tabular}
\end{center}
\caption{
Static-light decay constant ratios and the quenched continuum values. 
Values are given using $r_0=0.49(1)$ fm to set the physical $m_s$ 
point.
}
\label{tab:decayratio}
\end{table}

\section{Preliminary results for kinetic corrections}

\begin{figure}
\begin{center}
\resizebox{0.35\textwidth}{0.31\textwidth}{
\includegraphics[clip]{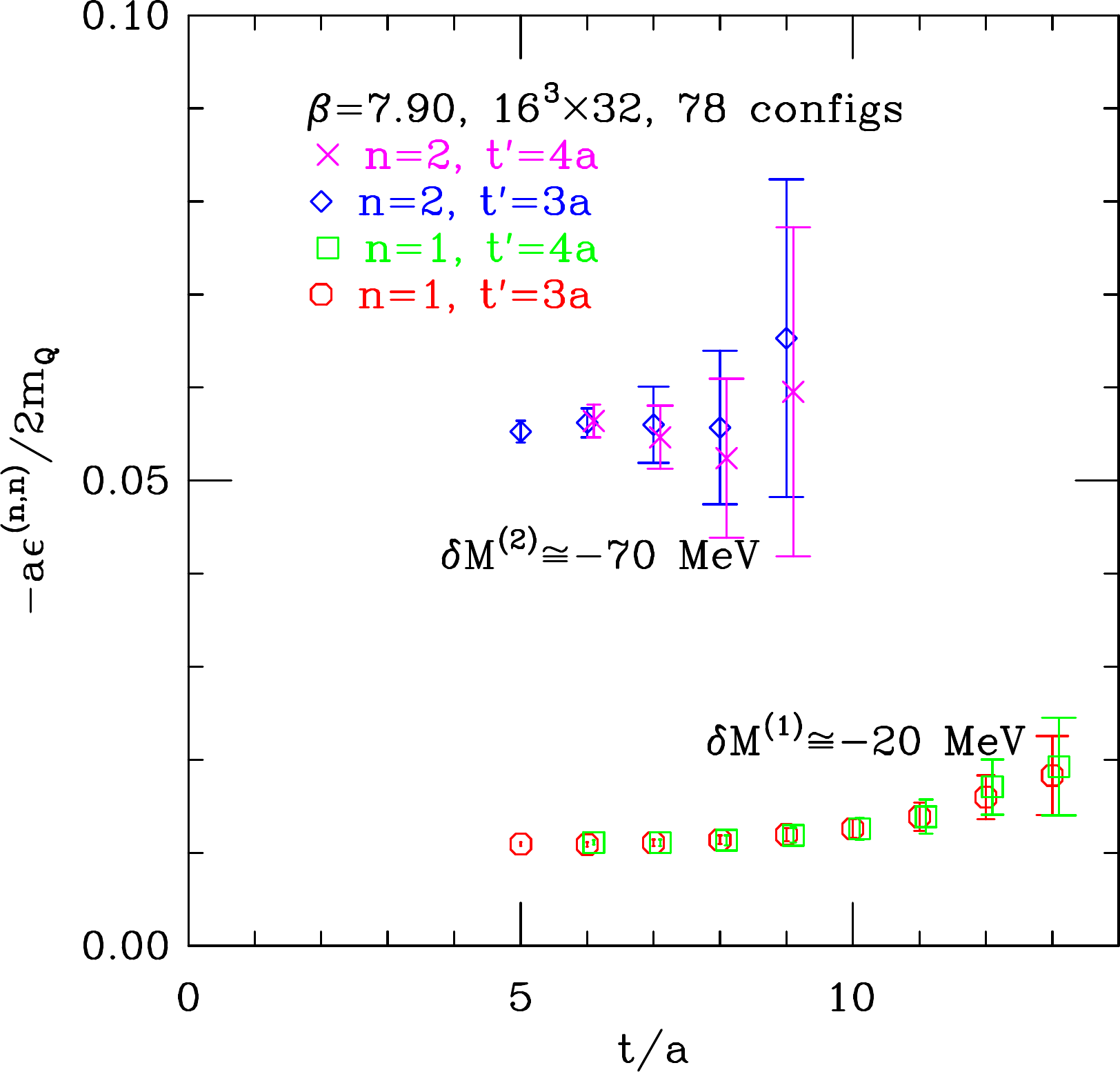}
}
\end{center}
\caption{
Preliminary results for the kinetic corrections. For the b quark mass we use $m_b \approx 4.2$ GeV and assume $Z\approx1$
for the renormalization constant.
}
\label{fig:kinetcorr}
\end{figure}

Finally we present also some preliminary results for kinetic corrections obtained 
via the generalization of the variational method described in Section \ref{sect:kincorr}. 
Figure \ref{fig:kinetcorr} shows first results for the $\mathcal{O}(1/m_b)$ corrections to the splitting 
$2S-1S$, obtained from $\epsilon^{(1,1)}$ and $\epsilon^{(2,2)}$. These are of course only rough estimates,
where we have assumed that the renormalization factor of the operator $\bar Q \vec D^2 Q$ is close to unity, since
we use a tadpole improved version of the operator $\vec D^2$. We also assume that the power-law divergent part of that
operator is small ($\epsilon^{(0,0)} \approx 0$) at this lattice spacing.
As mass of the b quark we took a value of 4.2 GeV as an input.

\section{Summary}

In our studies, we use the variational method to extract not only the masses of ground 
and excited states but also the couplings. In addition, we generalize this method such 
that it allows also for extraction of three point functions of excited states.

We successfully isolate excited static-light mesons via the variational method on a large 
number of lattices, including $2S$, $3S$, $1P$, $2P$, $1D$ states. In general our results 
show quite good agreement with the experimental values, where known. We perform also an 
extensive study for the static-light baryons. Also here we find a number of states.

In addition to the spectrum, we try to determine decay constants of the ground and excited 
static-light mesons. So far we can only report ratios of the decay constants since the 
necessary renormalization constants are not determined yet for the particular actions which 
we use.

As a next step we go beyond the static approximation by including kinetic corrections for the
b quark. We treat these corrections as current insertions and develop a way to apply the
variational method to three-point functions involving excited states. 
First results have been presented, assuming a mild renormalization of the inserted current operator
and using $m_b \approx 4.2$ GeV.

\acknowledgments

We would like to thank Rainer Sommer, Marc Wagner, and Georg von Hippel for interesting discussions 
at the lattice conference. 
This work is supported by GSI. 
The work of T.~B. is supported by the NSF (NSF-PHY-0555243) and the DOE (DE-FC02-01ER41183).
The work of M.~L. is supported by the DK W1203-N08 of the ``Fonds zur F\"orderung der wissenschaftlichen Forschung in \"Osterreich''.

\end{document}